\documentclass[ACS,STIX1COL]{WileyNJD-v2}

\articletype{Article Type}%

\usepackage{amsmath}    % need for subequations
\usepackage{amsthm}
\usepackage{amsfonts}
\usepackage{amssymb}
\usepackage{dsfont}
\usepackage{sidecap}
\sidecaptionvpos{figure}{t}

\received{xx August 2019}
\revised{xx August 2019}
\accepted{xx August 2019}

\newsavebox{\astrutbox}
\sbox{\astrutbox}{\rule[-5pt]{0pt}{20pt}}

\newcommand{\aeq}{\begin{equation}}
\newcommand{\eeq}{\end{equation}}
\newcommand{\aeqn}{\begin{eqnarray}}
\newcommand{\eeqn}{\end{eqnarray}}
\newcommand{\aeqs}{\begin{equation*}}
\newcommand{\eeqs}{\end{equation*}}
\newcommand{\aeqns}{\begin{eqnarray*}}
\newcommand{\eeqns}{\end{eqnarray*}}
\newcommand{\vecb}[1]{{\bf #1}}

\newcommand{\vpar}{{v_\parallel}}

\newcommand{\fpar}[2]{\frac{\partial{#1}}{\partial{#2}}}

\def \d {\mathrm{d}}

\begin{document}

\title{Collisional gyrokinetic full-f particle-in-cell simulations on open field lines with PICLS}

\author[1]{Mathias Boesl*}

\author[1]{Andreas Bergmann}

\author[1]{Alberto Bottino}

\author[2]{Stephan Brunner}

\author[1]{David Coster}

\author[1]{Frank Jenko}

\authormark{BOESL \textsc{et al}}

\address[1]{\orgdiv{Max Planck Institut f\"ur Plasmaphysik}, \orgaddress{\state{ D-85748 Garching}, \country{Germany}}}

\address[2]{\orgdiv{Swiss Plasma Center}, \orgname{Ecole Polytechnique F\'ed\'erale de Lausanne}, \orgaddress{\country{Switzerland}}}

\corres{Email address for correspondence: mathias.boesl@ipp.mpg.de}

\presentaddress{Mathias Boesl, Boltzmannstrasse 2, 85748 Garching}

\abstract{Applying gyrokinetic simulations for theoretical turbulence and transport studies to the plasma edge and scrape-off layer (SOL) presents significant challenges. 
To in particular account for steep density and temperature gradients in the SOL, the ``full-f" code PICLS was developed.
PICLS is a gyrokinetic particle-in-cell (PIC) code and is based on an electrostatic model with a linearized field equation and uses kinetic electrons. 
In previously published results we were applying PICLS to the well-studied 1D parallel transport problem during an edge-localized mode (ELM) in the SOL without collisions. As an extension to this collision-less case and in preparation for 3D simulations, in this work a collisional model will be introduced. 
The implemented Lenard-Bernstein collision operator and its Langevin discretization will be shown. Conservation properties of the collision operator as well as a comparison of the collisional and non-collisional case will be discussed.}


\jnlcitation{\cname{%
\author{Boesl M.}, 
\author{Bergmann A.}, 
\author{Bottino A.}, 
\author{Brunner S.}, 
\author{Coster D.}, and 
\author{Jenko F.}} (\cyear{2019}), 
\ctitle{Collisional gyrokinetic full-f particle-in-cell simulations on open field lines with PICLS}, \cjournal{TBD}, \cvol{TBD}.}

\maketitle

\section{Introduction\label{intro}}
In the closed field line region of the plasma core, gyrokinetics has become the workhorse for turbulence simulations in the last decades, but their extension towards the plasma edge and SOL reveals additional challenges.
Therefore, in a previous work \cite{UNKNOWN}, we investigated the well-studied problem of parallel energy and particle transport caused by a transient Type I ELM in the SOL.
Heat pulse simulations with a single central source model were already studied with fully-kinetic PIC, continuum (Vlasov) and fluid codes and successfully benchmarked against experiment. \citep{Havlickova12} 
Therefore, in our previous work we also studied this problem and achieved good agreement with very recent gyrokinetic continuum code simulations \cite{Shi15_1D, ShiPhD, Pan16_1D} in the collision-less case that reproduced the results of the mentioned previous works \citep{Havlickova12}.\\
However, collisions are a key driver to transport particles across closed magnetic flux surfaces that would be confined otherwise. \cite{Ricci15}.
They cause plasma to diffuse from the confined region into the SOL and from there they are transported towards the device wall. \cite{Ricci15}
Due to lower temperature, collisionality is also higher in the SOL than in the core region.
Therefore, in this study we implement a Lenard-Bernstein (LB) collision operator in our newly developed PICLS code, which is designed to perform gyrokinetic SOL simulations. \cite{UNKNOWN} For the applied PIC model the operator is discretized via a Langevin approach. \cite{Vernay13coll}
We will show that the implemented LB collision operator conserves particle number, parallel momentum and energy and relaxes towards a Maxwellian. Additionally, we will repeat our previously studied 1D1V heat pulse problem in a modified 1D2V version --- with the magnetic moment $\mu$ as additional coordinate --- and compare the collision-less with the collisional case.\\
The electrostatic gyrokinetic equations implemented in PICLS for the 1D heat pulse problem are described in section \ref{phys_model}. 
In section \ref{LBcoll_op} the considered LB collision operator and its PIC discretization is introduced. Simulation results for collision operator testing and the 1D2V collisional heat pulse problem are shown in section \ref{coll_simu_resul} and \ref{1D_simu_resul}. Section \ref{conclusio} contains conclusions and an outlook.
\section{Physical model}\label{phys_model}
The equations implemented in PICLS are derived from a low-frequency and electrostatic gyrokinetic model with kinetic electrons. Due to the 1D ELM pulse problem we investigate here, finite-Larmor radius effects are not required. However, Larmor-radius effects and gyroaveraging are already implemented for future higher dimensional simulations. 
As of now, PICLS is purely based on slab geometry and for the 1D heat pulse problem just the 1D slab versions of the Euler-Lagrange eqs. for position $z$ and parallel velocity $\vpar$ are required. By choosing $B=\textrm{const}$ and parallel to the z-direction, for species $p$ these can be written as:
\aeqn
\dot{z}=\vpar\vecb{b}, \quad
\dot{\vpar }= - \frac{e_p}{m_p}\vecb{b}\cdot \nabla J_{p,0}\phi
\label{Slabeqmo1D}
\eeqn
with the gyroaveraging operator $J_{p,0}$. By introducing the shielding factor $s_\perp(z,t)=k_\perp^2(z) \epsilon_\perp(z,t)$, with $\epsilon_\perp= \sum_p\frac{n_{p,0} m_pc^2}{B^2}$, a simplified polarization equation can be obtained that only takes a single perpendicular wave number $k_\perp$ into account: \cite{ShiPhD, Pan16_1D}
\aeqn
s_\perp(z,t)(\phi-\langle\phi\rangle) = \sum_p \int \d W  e_pJ_{p,0} f.
\label{polareq1D_mod}
\eeqn
Where the flux-surface-averaged, dielectric-weighted potential
$\langle\phi\rangle = \frac{\int \d z s_\perp \phi}{\int \d z s_\perp}$ is used.
For more details on the derivation of the physical model and its numerical discretization, we refer to our previously published work. \cite{UNKNOWN}\\
A logical sheath model is implemented to model the effects of a Debye sheath, without actually having to resolve it. 
The setup of the implemented logical sheath is generally based on the model shown in Parker et al. \cite{parker93sheath}, which was developed for fully kinetic 1D2V PIC simulations. 
The same model was used for previous parallel heat flux studies with gyrokinetic 1D1V continuum-codes. \cite{Shi15_1D, Pan16_1D}
Here, the total parallel current $j_\parallel$ to the wall is set to $0$. 
This model mimics the physical effect of accelerating incident ions by the dropping sheath potential $\phi_\textrm{sh}$. 
For electrons however the velocity needs to be high enough to overcome the $\phi_\textrm{sh}$ drop at the boundary and slower electrons are reflected backwards.
With the wall potential $\phi_\textrm{w}$ ($\phi_\textrm{w}=0$ for a grounded wall), the electron cut-off velocity $v_\textrm{ce}$, which is the velocity of the slowest electron exiting the domain, determines $\phi_\textrm{sh}$ according to: 
\aeqn
	\delta\phi=\phi_\textrm{sh}-\phi_\textrm{w}=\frac{m}{2e}v_\textrm{ce}^2.
\label{phisheath}
\eeqn
\section{Lenard-Bernstein (LB) collision operator} \label{LBcoll_op}
To account for collisions in our model, the LB collision operator is implemented. 
It can be used in the presence of small-angle collisions and includes collision driven diffusion in velocity space, which causes the distribution function to relax towards a Maxwellian. The results of a Landau operator are recovered in the limit of infrequent collisions. \cite{Lenard58} However, in the simplified LB op. the evaluation of Rosenbluth potentials is avoided.
The operator contains pitch-angle scattering \& conserves particle number, momentum, and energy analytically. 
It assumes long wavelength, i. e. ignores finite-Larmor-radius (FLR) corrections. 
The LB collision operator acting on the full-f model for species $p$ and $p'$ is written as:
\aeqn \label{coll_op}
C_p[f_p] &=& \sum_{p'}C_{pp'}[f_p]
 = \sum_{p'}\nu_{pp'}\fpar{}{\vecb{v}}\cdot \left[(\vecb{v}- \vecb{u_{\parallel,p'}})f_p + v^2_{T,pp'} \fpar{f_p}{\vecb{v}} \right]\nonumber\\
  &=& \sum_{p'}\nu_{pp'}\left(
  \fpar{}{\vpar} \left[(\vpar- u_{\parallel,p'})f_p + v^2_{T,pp'} \fpar{f_p}{\vpar} \right] +  \fpar{}{\mu} \left[2\mu f_p + 2 \frac{m_p v^2_{T,pp'}}{B}\mu \fpar{f_p}{\mu} \right]\right)  
\eeqn
with the definitions:
\itemsep0em
\begin{center}
\begin{tabularx}{\textwidth}{XXX}
\begin{equation}\label{upar_eq}
    u_\parallel = \frac{\int \d^3v v_\parallel f_p}{n_p}
\end{equation}
    &
\begin{equation}\label{vtherm_eq}
    m_p v^2_{T,pp'}=\frac{\int\d^3v m_p (\vecb{v}-\vecb{u_{p'}})^2 f_p}{3n_p}
\end{equation}
	&
\begin{equation}\label{dens_eq}
    n_p=\int \d^3 v f_p.
\end{equation}
\end{tabularx}
\end{center}
Additionally, for the collision frequencies of self-species collisions standard expressions can be used that are defined as $\nu_\textrm{ee} = \frac{4\sqrt{2\pi}n_\textrm{e} \lambda e^4}{3\sqrt{m_\textrm{e}}T_\textrm{e}^{3/2}}$ and $\nu_\textrm{ii} = \frac{4\sqrt{\pi}n_\textrm{i} \lambda e^4}{3\sqrt{m_\textrm{i}}T_\textrm{i}^{3/2}}$. Here $\lambda$ stands for the Coulomb logarithm $\lambda = 6.6 - 0.5 \ln(n_0/10^{20})+1.5\ln T_{e0}$, where $n_0$ is expressed in $m^{-3}$ and $T_{\textrm{e}0}$ in eV.  
For electron-ion collisions the LB collision operator is also used for simplicity with the collision frequency $\nu_\textrm{ei}=\nu_\textrm{ee}/1.96$, which approximately accounts for the plasma's parallel conductivity coefficient. Ion-electron collisions are neglected, since $\nu_\textrm{ie}$ is much smaller than the ion-ion term $(\nu_\textrm{ie}/\nu_\textrm{ii}\sim\sqrt{m_\textrm{e}/m_\textrm{i}})$, as also done in gyrokinetic continuum code studies in Gkeyll and GENE. \cite{Shi17,Pan18} 
The drag coefficient $\Gamma$ and diffusion coefficient $D$ can be extracted from equation \eqref{coll_op}:
\aeqn
 \Gamma = - \nu_{pp'}(\vecb{v}- \vecb{u_{\parallel,p'}}),\quad
 D = \nu_{pp'} v^2_{T,pp'}.\label{draganddiff_term}
\eeqn
To discretize the LB collision operator for our PIC approach, we use the so-called Langevin approach as explained in Vernay et al. \citep{Vernay13coll} 
Applying this approach, the position in phase space $x_n(t)$ of marker $n$ at time $t$, is given by its previous position $x_n(t-\Delta t)$ at time step $t - \Delta t$, according to:
\aeqn
 \Delta x_n = x_n(t)-x_n(t-\Delta t) = \langle \Delta x\rangle + \mathcal{R} \sqrt{\langle \Delta x \Delta x\rangle}=\Gamma \Delta t + \mathcal{R} \sqrt{2D\Delta t},
\label{langevin_delta}
\eeqn
where $\mathcal{R}$ is a random number sampled from a PDF of average $0$ and variance $1$. 
To ensure that $\xi_\textrm{out}=\vpar/v\in [-1,1]$, one temporarily expands the 2D gyrokinetic velocity space to 3D. \cite{Vernay13coll}
Using the drag and diffusion coefficients \eqref{draganddiff_term} in \eqref{langevin_delta}, we thus achieve the velocity change in the $(v_x,v_y,v_z)$ space:
\aeqn
 \Delta v_x = -\nu_{pp'}v_{\perp,in}\Delta t + v_{T,pp'}\sqrt{2\nu_{pp'} \Delta t}\mathcal{R}_1,\;
 \Delta v_y = v_{T,pp'}\sqrt{2\nu_{pp'} \Delta t}\mathcal{R}_2, \;
 \Delta v_z = -\nu_{pp'}(v_{\parallel,\textrm{in}}-u_{\parallel,p'})\Delta t + v_{T,pp'}\sqrt{2\nu_{pp'} \Delta t}\mathcal{R}_3, \label{delta_vs}
 \eeqn
with $v_\perp=\sqrt{2B(\vecb{R})\mu/m}$ and the independent random numbers $\mathcal{R}_1$, $\mathcal{R}_2$ and $\mathcal{R}_3$.  To achieve the velocity values for the out-going marker after the collision operation, one has to reverse transform the coordinates back to the 2D velocity space:
\aeqn
\vpar_{,\textrm{out}} = \vpar_{,\textrm{in}} + \Delta v_z,\quad
v_{\perp,\textrm{out}} = \sqrt{(v_{\perp,\textrm{in}}+\Delta v_x)^2+\Delta v_y^2}.
\eeqn
For the collision operator implementation, conservation of particle number, parallel momentum $\sim\langle \vpar \rangle$ and kinetic energy $\sim\langle v^2 \rangle$ is decisive. Analytically, the LB operator conserves these quantities for infinitely small time steps and an infinite number of particles. However with finite values, corrections can be introduced to ensure that conservation relations hold up to round-off.  
Since PICLS is based on a full-f model, particle number is intrinsically conserved.
For $\langle \vpar \rangle$ and $\langle v^2 \rangle$, the idea however is to regard $u_\parallel$ and $v_T$ as free parameters, which are determined in order to ensure conservation of moments.
Here, we relax the condition of finite particle number, and only ensure that the conservation holds on average over the statistics of Langevin kicks, but $\Delta t$ is kept finite. This equation is implemented in PICLS and for our use shows good conservation of moments, as shown in section \ref{coll_simu_resul}.
For the parallel momentum, we set the change of the average parallel velocity to zero, which corresponds to $\Delta v_z$ from equation \eqref{delta_vs}, since $\vpar$ lies in the $z$-direction. We sum over all marker weights within a configuration space bin, to achieve:
\aeqn
0=\sum_{n=1}^{N}w_n\Delta v_{\parallel,n} = \sum_{n=1}^{N}w_n\Delta v_{z,n} 
= \sum_{n=1}^{N} w_n \left[ -\nu(v_{\parallel,in,n}-u_{\parallel})\Delta t + v_{T}\sqrt{2\nu \Delta t}\mathcal{R}_{3,n}\right],
\eeqn
with $N$ the total number of markers within the bin. Using the relation $\langle\mathcal{R}_{3,n}\rangle=0$, which comes from our choice of the PDF, the second term drops and we can achieve a relation for $\langle u_\parallel\rangle$ to conserve parallel momentum:
\aeqn
\langle u_\parallel\rangle 
= \sum_{n=1}^{N} w_n v_{\parallel,in,n}/\sum_{n=1}^{N} w_n,
\label{upar_relation}
\eeqn
which is exactly the obvious PIC discretization relation for \eqref{upar_eq}. Thus, no correction for $u_\parallel$ is required to conserve parallel momentum. 
The next step is to derive a relation for $v_T$ in order to conserve energy. For this, the derived relation for $u_\parallel$ \eqref{upar_relation} can be used. On average over all particles the following relation for the total change of kinetic energy must hold:
\aeqn
0= \vecb{v}_\textrm{out}^2-\vecb{v}_\textrm{in}^2= (\vecb{v}+\Delta\vecb{v})^2 - \vecb{v}^2 = 2\vecb{v}\Delta\vecb{v}+(\Delta\vecb{v})^2.
\label{kinEn_relation}
\eeqn
Writing the explicit expression for $\Delta\vecb{v}$ that follows from eq. \eqref{langevin_delta} and summing over the markers yields:
\aeqn
0&=&\sum_{n=1}^{N} w_n \Big\{ 2(\vecb{v}_{\textrm{in},n}- \vecb{e}_z u_{\parallel}) \cdot \left[ -\nu \Delta t(\vecb{v}_{\textrm{in},n}- \vecb{e}_z u_{\parallel}) + v_T \sqrt{2\nu \Delta t} \vecb{\mathcal{R}}_n\right]
+ \left[ -\nu \Delta t(\vecb{v}_{\textrm{in},n}- \vecb{e}_z u_{\parallel}) + v_T \sqrt{2\nu \Delta t} \vecb{\mathcal{R}}_n\right]^2 \Big\}\\
&=& \sum_{n=1}^{N} w_n \Big\{ -\nu \Delta t(2-\nu \Delta t) (\vecb{v}_{\textrm{in},n}- \vecb{e}_z u_{\parallel} )^2
+ 2v_T \sqrt{2\nu \Delta t}(1-\nu \Delta t) (\vecb{v}_{\textrm{in},n}- \vecb{e}_z u_{\parallel} )\vecb{\mathcal{R}}_n + 2v_T^2 \nu \Delta t \vecb{\mathcal{R}}_n^2
\Big\}.
\eeqn
Invoking the properties of the PDF for the random numbers $\langle \vecb{\mathcal{R}}_n \rangle=0$ and $\langle \vecb{\mathcal{R}}_n^2 \rangle=\langle \mathcal{R}_{1,n}^2 \rangle+\langle \mathcal{R}_{2,n}^2 \rangle+\langle \mathcal{R}_{3,n}^2 \rangle=3$ one obtains:
\aeqn
3(\sum_{n=1}^{N} w_n)2\nu \Delta t \langle v_T^2\rangle = \nu \Delta (2- \nu \Delta t) \sum_{n=1}^{N} w_n(\vecb{v}_{\textrm{in},n}- \vecb{e}_z u_{\parallel} )^2.
\label{vT_calc}
\eeqn
A relation for $\langle v_T^2\rangle$ to conserve kinetic energy can directly be derived from eq. \eqref{vT_calc}:
\aeqn
\langle v_T^2\rangle = (1- \nu \Delta t/2)\frac{\sum_{n=1}^{N} w_n(\vecb{v}_{\textrm{in},n}- \vecb{e}_z u_{\parallel} )^2}{3\sum_{n=1}^{N} w_n}.
\label{vT_relation}
\eeqn
This relation appears as the appropriate PIC discretization of \eqref{vtherm_eq}. Note the correction factor $(1- \nu \Delta t/2)$, which is required to achieve conservation of kinetic energy for finite time steps.
\section{Simulation results: 1D2V collision operator testing}\label{coll_simu_resul}
\begin{SCfigure}
\centering
\includegraphics[width=0.7\textwidth]{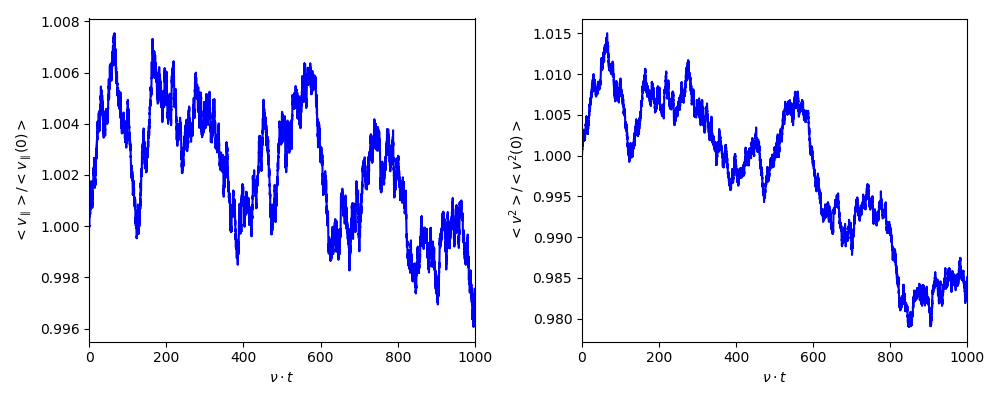}
\caption{
Time evolution of $\langle \vpar(t)\rangle/\langle \vpar(0)\rangle$ and $\langle v(t)^2\rangle/\langle v(0)^2\rangle$ for simulations with self-consistent calculation of $\vecb{u_{\parallel}}$ and $v_T$ plotted for a total simulation time (in $\nu\cdot t$). Parallel momentum and kinetic energy are mostly conserved with max. $2\%$ deviation from initial values.
}
\label{colltest_conserve_moments}
\end{SCfigure}

To decrease complexity and test against analytic functions, we choose a single species $p$ subject to self-species collisions and set $\vecb{u_{\parallel}}=0$, $v_T=\textrm{const}$ and $\nu=\textrm{const}$. With this setting, the evolution eq. for the distribution $f$ is given by:
\aeqn 
\fpar{}{t}f = C_p[f] = \nu\fpar{}{\vecb{v}}\cdot \left[\vecb{v}f + v^2_{T} \fpar{f}{\vecb{v}} \right].
\eeqn
With the definitions for density $n = \int \d^3v f$, average velociy $\langle \vecb{v}\rangle= \int \d^3v \vecb{v} f/n $ and kinetic energy $\langle v^2\rangle= \int \d^3v v^2 f/n$, analytic expressions can be derived for the time evolution of these quantities and compared with the numerical simulations.
For $\langle \vecb{v}\rangle$ and $\langle v^2\rangle$ the following exponentially decaying functions can be found as solutions:
\aeqn 
\langle \vecb{v}\rangle(t)=\langle \vecb{v}\rangle(t=0)e^{-\nu t}, \quad
\langle v^2\rangle(t)=3v^2_{T}+\left[\langle v^2\rangle(t=0) -3v^2_{T}\right] e^{-2\nu t}.
\label{vvec_v2_func}
\eeqn
By choosing an arbitrary initial velocity distribution, which has to relax according to eqs. \eqref{vvec_v2_func}, we can construct a first test case for the implemented collision operator. Performing this test case shows that the implemented operator is able to reproduce the analytic results of eqs. \eqref{vvec_v2_func} and the marker distribution relaxes to a Maxwellian with the considered values for $\vecb{u_{\parallel}}$ and $v_T$ in the equilibrated state.\\
%In figure \ref{colltest_basic_expdecay}, the time evolution of $\langle \vpar\rangle$ and $\langle v^2\rangle$ is plotted in non-normalized units. Together with the simulation results, also the analytic solutions \eqref{vvec_func} and \eqref{v2_func} are plotted and show very good agreement. The implemented operator apparently is able to reproduce the analytic results in this simplified test case.
%\begin{SCfigure}\label{colltest_basic_expdecay}
%\centering
%\includegraphics[width=0.6\textwidth]{../../solpictex/FIGURES/colltest_basic_expdecay.png}
%\caption{
%Time evolution of $\langle \vpar\rangle$ and $\langle v^2\rangle$ in non-normalized units plotted per time step until a steady state is reached. The simulation results (blue) show very good agreement with the analytic solutions (green) from equations \eqref{vvec_func} and \eqref{v2_func}.
%}
%\end{SCfigure}
%We just want to mention, that the marker distribution in $\vpar$ reaches a Maxwellian with the defined values for $\vecb{u_{\parallel}}$ and $v_T$ in the equilibrated state. In $\xi=\vpar/v$ the distribution reaches an equilibrated state with equal distribution of parallel and perpendicular velocity components. 
As a second test case, the more general form of the collision op. as in \eqref{LBcoll_op} is considered, which calculates $\vecb{u_{\parallel}}$ and $v_T$ at each time step acc. to \eqref{upar_relation} and \eqref{vT_relation}. Here, the previously mentioned correction factor $(1- \nu \Delta t/2)$ for estimating $v_T$ is implemented.
For conservation tests, one single species and thus only self-species collisions are used. An arbitrary initial particle distribution should relax towards a Maxwellian in $\vpar$ and conserve $\langle \vecb{v} \rangle$ and $\langle v^2\rangle$. The number of particles is automatically conserved, due to the chosen full-f model. 
Figure \ref{colltest_conserve_moments}shows the time evolution of $\langle \vpar(t)\rangle/\langle \vpar(0)\rangle$ and $\langle v(0)^2\rangle/\langle v(0)^2\rangle$ for an exemplary simulation to highlight the changes of parallel momentum and kinetic energy. 
Despite choosing an approx. 10--20 times lower particle resolution per bin than in the 1D heat pulse simulations in section \ref{1D_simu_resul}, in figure \ref{colltest_conserve_moments}$\langle \vpar \rangle$ and $\langle v^2 \rangle$ are mostly conserved with only a variation of $<2\%$ around its initial value. 
By increasing the number of particles, the deviations even get smaller.  
Again, an arbitrary initial marker distribution in $\vpar$ relaxes to a Maxwellian, but this time its maximum remains at the initialized $\vecb{u_{\parallel}}$. Since the parallel momentum conservation property holds, the particle velocities remain distributed around their initial $\vecb{u_{\parallel}}$. 
\section{Simulation results: Collisional 1D2V heat pulse}\label{1D_simu_resul}
\begin{SCfigure}
\centering
\includegraphics[width=0.79\textwidth]{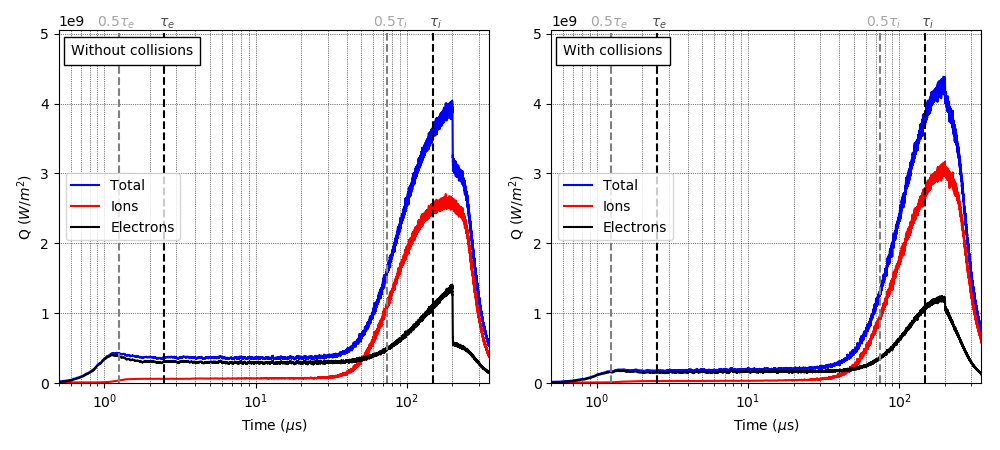}
\caption{
Comparison of the evolution of ion (red), electron (black) and total (blue) heat flux in the 1D2V case, according to eq. \eqref{coll_heatflux}, with and without same-species Lenard-Bernstein collisions. Thermal ion and electron transit times $\tau_\textrm{e}=2.5\mu s$ and $\tau_\textrm{i}=149\mu s$ are indicated by black vertical lines ($0.5\tau_\textrm{e}$ and $0.5\tau_\textrm{i}$ are indicated by grey lines).
}
\label{heatflux_coll_comparison}
\end{SCfigure}

\begin{SCfigure}
\centering
\includegraphics[width=0.6\textwidth]{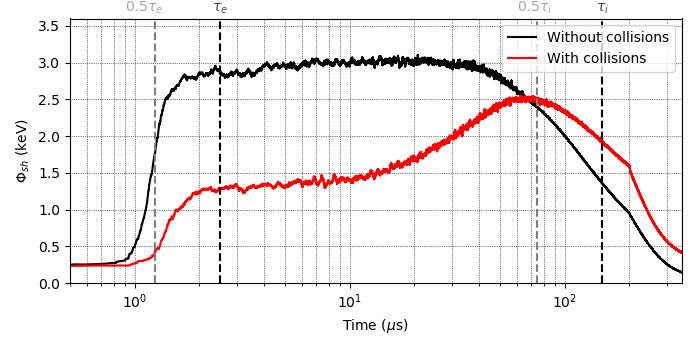}
\caption{
Comparison of the time-dependent evolution of the sheath potential at the right boundary in the 1D2V case with and without same-species Lenard-Bernstein collisions. The vertical black and grey lines are at the same position as in figure \ref{heatflux_coll_comparison}on page \pageref{heatflux_coll_comparison}.
}
\label{sheathpot_coll_comparison}
\end{SCfigure}

In our previous publication \cite{UNKNOWN}, we performed simulations on a 1D heat pulse in the scrape-off layer without collisions. There, a high energy particle source acted as an ELM heat pulse and injected particles for $200\mu s$. Within this work, we again want to use the same setup. However, in view of realistic SOL simulations, particle collisions are introduced. 
We therefore introduce the magnetic moment $\mu$ as a second velocity component and use the LB collision operator, described in section \ref{LBcoll_op}. Inter-species collisions between electrons and ions are neglected similar to the work done by Shi et al. \cite{Shi15_1D}\\ 
By introducing $\mu$, also the fixed $T_\perp$ from our previous work\cite{UNKNOWN} now can be changed over time by the collision operator. For the heat pulse source, the perpendicular temperature however is kept constant at $T_\textrm{ped}$, even after the ELM heat pulse ends. For the parallel temperature, the same setup as described in Boesl et. al. \cite{UNKNOWN} will be used. 
With $\mu$ as additional velocity component, the equation for the parallel heat flux can be written as:
\aeqn
Q_p = \int_{v_{\textrm{c},p}}^{\infty}f_p \vpar\left(\frac{1}{2}m_p \vpar^2 + \mu B\right)\d \vpar + q_p \phi_\textrm{sh} \int_{v_{\textrm{c},p}}^{\infty} f_p \vpar\d \vpar.
\label{coll_heatflux}
\eeqn
Figure \ref{heatflux_coll_comparison}compares the heat flux on the divertor wall for non-collisional and collisional 1D2V simulations. We want to mention that the values in the 1D2V collision-less case differ from the 1D1V simulations of our previous work \cite{UNKNOWN}, due to the source applied for the $\mu$ initialization. 
The first differences we notice between both plots is a lower initial heat flux before $\sim 0.5 \tau_\textrm{i}$ for the collisional case of about $50\%$ of the non-collisional case. 
Once the suprathermal ions hit the wall, the ion heat flux in the collisional case rises even higher than in the non-collisional case. However, for the electrons a slight decrease in the maximum heat flux is visible. 
For the total heat flux an $\sim 8\%$ higher maximal value ($4.04\cdot 10^9W/m^2$ vs. $4.38\cdot 10^9W/m^2$) thus is reached in the collisional case.
Further investigating the heat flux in the collisional (non-collisional) case reveals more differences. 
The share of the total heat flux over time deposited before the peak at $200\mu s$ is $55\%$ ($61\%$) and for the total heat flux deposited by ions vs. electrons we get shares of $74\%$ vs. $26\%$ ($72\%$ vs. $28\%$). The total heat flux deposited over time in the collisional case is $9\%$ higher than in the non-collisional. This clearly shows, that the collisions introduced lead to an increase in the ion heat flux and in total over time lead to a higher heat flux on the wall. 
This largely depends on the increased particle flux, but to better understand the heat flux evolution, in figure \ref{sheathpot_coll_comparison}a comparison of the sheath potential $\phi_\textrm{sh}$ with and without collisions is shown. First, $\phi_\textrm{sh}$ in both cases is determined by the cold initial distribution. But at $\sim0.5-1$ $\tau_\textrm{e}$ both curves increase rapidly, due to arriving suprathermal electrons from the ELM source. 
In the collision-less case, $\phi_\textrm{sh}$ immediately rises to $\sim3$keV, where it stays mainly constant until the arrival of suprathermal ions at $\sim0.5\tau_\textrm{i}$. On the other hand, in the collisional case
$\phi_\textrm{sh}$ rises quickly to $\sim1.5$keV and then gradually increases up to its maximum of $\sim2.5$keV at $\tau_\textrm{i}$. This is an indicator, that the collision operator decreases the high-$\vpar$ tail of the velocity distribution, as a result of drag on ions or thermalisation through self-collisions.
In both cases the $\phi_\textrm{sh}$ decelerates and reflects electrons at the wall. The majority of electrons are prevented to leave the domain and thus the increase of the electron flux is stopped at $\sim0.5\tau_\textrm{e}$. The sheath potential decreases steadily after the inflow of suprathermal ions and allows an increase of both the ion and electron flux. However, $\phi_\textrm{sh}$ remains higher for the collisional case after $\sim0.5\tau_\textrm{i}$. This again can be described by the collision operator, which is able to replenish high-$\vpar$ electrons through pitch-angle scattering.
\section{Conclusions}\label{conclusio}
Different from our previous publication \cite{UNKNOWN}, in the current work a Lenard-Bernstein collision operator for same-species collisions was implemented in the gyrokinetic PIC code PICLS. The operator's PIC discretization and conservation properties were discussed and a correction term to conserve energy up to round-off for finite time steps and infinite number of particles was derived. Using this correction for finite number of particles a very good energy conservation could still be shown.
Following our previous work \cite{UNKNOWN}, with the new collisional model, 1D2V heat pulse simulations were performed and compared with non-collisional results. 
Collisions had a significant effect on the heat flux deposited on the sheath, which in total increases by $9\%$. And also the sheath potential $\phi_\textrm{sh}$ undergoes a deferred increase and lower maximum of $\sim2.5$keV (compared to $\sim3.0$keV), due to collisional effects. 
Having a working collisional model implemented, PICLS is now prepared to extend to higher spatial dimension for future simulations. 
\section*{Acknowledgments}
The authors would like to thank L. Villard, N. Ohana and E. Lanti from the Swiss Plasma Center (Lausanne) for their help. Numerical simulations were performed on the Marconi supercomputer within the framework of the PICLS project. This work has also been carried out within the framework of the EUROfusion Consortium and has received funding from the Euratom research \& training program 2014-2018 and 2019-2020 under grant agreement No 633053, for the EF WP32-ENR-MPG-04 (2019/2020) project ``MAGYK". The views and opinions expressed herein do not necessarily reflect those of the European Commission.

{\footnotesize
\setlength{\bibsep}{0pt plus 0.3ex}
\bibliography{mathias}%

\providecommand{\url}[1]{\texttt{#1}}
\providecommand{\urlprefix}{}
\providecommand{\foreignlanguage}[2]{#2}
\providecommand{\Capitalize}[1]{\uppercase{#1}}
\providecommand{\capitalize}[1]{\expandafter\Capitalize#1}
\providecommand{\bibliographycite}[1]{\cite{#1}}
\providecommand{\bbland}{and}
\providecommand{\bblchap}{chap.}
\providecommand{\bblchapter}{chapter}
\providecommand{\bbletal}{et~al.}
\providecommand{\bbleditors}{editors}
\providecommand{\bbleds}{eds: }
\providecommand{\bbleditor}{editor}
\providecommand{\bbled}{ed.}
\providecommand{\bbledition}{edition}
\providecommand{\bbledn}{ed.}
\providecommand{\bbleidp}{page}
\providecommand{\bbleidpp}{pages}
\providecommand{\bblerratum}{erratum}
\providecommand{\bblin}{in}
\providecommand{\bblmthesis}{Master's thesis}
\providecommand{\bblno}{no.}
\providecommand{\bblnumber}{number}
\providecommand{\bblof}{of}
\providecommand{\bblpage}{page}
\providecommand{\bblpages}{pages}
\providecommand{\bblp}{p}
\providecommand{\bblphdthesis}{Ph.D. thesis}
\providecommand{\bblpp}{pp}
\providecommand{\bbltechrep}{}
\providecommand{\bbltechreport}{Technical Report}
\providecommand{\bblvolume}{volume}
\providecommand{\bblvol}{Vol.}
\providecommand{\bbljan}{January}
\providecommand{\bblfeb}{February}
\providecommand{\bblmar}{March}
\providecommand{\bblapr}{April}
\providecommand{\bblmay}{May}
\providecommand{\bbljun}{June}
\providecommand{\bbljul}{July}
\providecommand{\bblaug}{August}
\providecommand{\bblsep}{September}
\providecommand{\bbloct}{October}
\providecommand{\bblnov}{November}
\providecommand{\bbldec}{December}
\providecommand{\bblfirst}{First}
\providecommand{\bblfirsto}{1st}
\providecommand{\bblsecond}{Second}
\providecommand{\bblsecondo}{2nd}
\providecommand{\bblthird}{Third}
\providecommand{\bblthirdo}{3rd}
\providecommand{\bblfourth}{Fourth}
\providecommand{\bblfourtho}{4th}
\providecommand{\bblfifth}{Fifth}
\providecommand{\bblfiftho}{5th}
\providecommand{\bblst}{st}
\providecommand{\bblnd}{nd}
\providecommand{\bblrd}{rd}
\providecommand{\bblth}{th}
\begin{thebibliography}{10}

\bibitem{UNKNOWN}
M~Boesl, A~Bergmann, A~Bottino, D~Coster, E~Lanti, N~Ohana, F~Jenko, {\it arXiv
  preprint arXiv:1908.00318} \textbf{2019}.

\bibitem{Havlickova12}
E~Havl{\'\i}{\v{c}}kov{\'a}, Wojtek Fundamenski, David Tskhakaya, Giovanni
  Manfredi, Derek Moulton, {\it Plasma Physics and Controlled Fusion}
  \textbf{2012}, {\it 54} (4), 045002.

\bibitem{Shi15_1D}
EL~Shi, AH~Hakim, GW~Hammett, {\it Physics of Plasmas} \textbf{2015}, {\it 22}
  (2), 022504.

\bibitem{ShiPhD}
EL~Shi, {\it arXiv preprint arXiv:1708.07283} \textbf{2017}.

\bibitem{Pan16_1D}
Q~Pan, D~Told, F~Jenko, {\it Physics of Plasmas} \textbf{2016}, {\it 23} (10),
  102302.

\bibitem{Ricci15}
P~Ricci, {\it Journal of Plasma Physics} \textbf{2015}, {\it 81} (2),
  435810202.

\bibitem{Vernay13coll}
T~Vernay, Collisions in Global Gyrokinetic Simulations of Tokamak Plasmas using
  the Delta-f Particle-In-Cell Approach, {\it \bbltechrep{}}, EPFL,
  \textbf{2013}.

\bibitem{parker93sheath}
SE~Parker, RJ~Procassini, CK~Birdsall, BI~Cohen, {\it Journal of Computational
  Physics} \textbf{1993}, {\it 104} (1), 41--49.

\bibitem{Lenard58}
A~Lenard, IB~Bernstein, {\it Physical Review} \textbf{1958}, {\it 112} (5),
  1456.

\bibitem{Shi17}
EL~Shi, GW~Hammett, T~Stoltzfus-Dueck, A~Hakim, {\it Journal of Plasma Physics}
  \textbf{2017}, {\it 83} (3).

\bibitem{Pan18}
Q~Pan, D~Told, EL~Shi, GW~Hammett, F~Jenko, {\it Physics of Plasmas}
  \textbf{2018}, {\it 25} (6), 062303.

\end{thebibliography}
}

%\section*{Author Biography}

%\begin{biography}{\includegraphics[width=60pt,height=70pt,draft]{empty}}{\textbf{Author Name.} This is sample author biography text this is sample author biography text this is sample author biography text this is sample author biography text this is sample author biography text this is sample author biography text this is sample author biography text this is sample author biography text this is sample author biography text this is sample author biography text.}
%\end{biography}

\end{document}